\documentclass[12pt,oneside,english,12pt]{amsart}
\pdfoutput=1
\usepackage[T1]{fontenc}
\usepackage[latin9]{inputenc}
\usepackage{textcomp}
\usepackage{amstext}
\usepackage{amsthm}
\usepackage{amssymb}
\usepackage{graphicx}
\usepackage{hyperref}

\makeatletter

\providecommand{\tabularnewline}{\\}

\numberwithin{equation}{section}
\numberwithin{figure}{section}

\usepackage{amsfonts}

\makeatother

\usepackage{babel}
\begin{document}

\title{Critical behaviour of a non-local $\phi^{4}$ field theory and asymptotic freedom }

\author{R. Trinchero}
\begin{abstract}
The critical behaviour of a non-local scalar field theory is studied.
This theory has a non-local kinetic term which involves a real power
$1-2\alpha$ of the Laplacian. The interaction term is the usual local
$\phi^{4}$ interaction. The lowest order Feynman diagrams corresponding
to coupling constant renormalization, mass renormalization and field
renormalization are computed. Particular features appearing in the
renormalization of these non-local theory that differ from the case
of local theories are studied. The previous calculations lead to the
perturbative computation of the coupling constant beta function and
critical exponents $\nu$ and $\eta$. In four dimensions for $\alpha<0$
this beta function presents asymptotic freedom in the UV. This is
remarcable since no non-abelian vector fields are included. However
this comes at the expense of loosing reflection positivity.
\end{abstract}

\date{17/04/2018}
\maketitle

\section{Introduction}

The computation of critical exponents for the $3$-dimensional Ising
model using the $\epsilon$-expansion provides a concrete example
of the relevance of the renormalization group ideas\cite{Wilson:1971dc}\cite{Wilson:1973jj}.
This is done by considering a self -interacting $\phi^{4}$ theory
in $d=4-\epsilon$ dimensions, where $\epsilon$ is allowed to take
real values. This procedure led to a qualitative understanding of
the $3$-dimensional Ising model physics and to predictions for critical
exponents in reasonable agreement with the exact values. The results
for the theory in $d$-dimensions are obtained by computing the theory
in a integer number $n$ of dimensions and then replacing $n$ by
$d$.

The renormalization group consists in the study of the evolution of
a system under scale transformations. This system involves all possible
interactions of any range for all kinds of dynamical variables. Different
physical systems correspond to the study of particular fixed points
in this huge space of couplings. This paper studies a particular example
of system described near the corresponding fix point by a non-local
field theory. The use of non-local field theories in the description
of critical phenomena is not new\cite{Fateev:1985mm},\cite{Reddy2014Auth},\cite{Paulos:2015jfa},
\cite{Egolf2017TheMF}. Such models appear in statistical systems
with long range interactions. In this paper the critical behaviour
of a non-local field theory is studied. This non-local theory is motivated
by an alternative approach to non-integer dimensional spaces(NIDS)\cite{Trinchero:2012zn}.
Free scalar theories on these spaces has been studied in this last
reference.  This theory has been employed to compute loop corrections
and compare the results with dimensional regularization\cite{bollini1972dimensional}\cite{tHooft:1972fi},
showing that the structure of singularities is the same as in dimensional
regularization. In addition, the fullfilment or not of the requirement
of reflection positivity for the corresponding Euclidean field theory
has beeen considered\cite{Trinchero:2017weu}. There, it is shown
that for negative values of the non-integer power mentioned above
the theory fullfills reflection positivity. This means that the corresponding
theory in Minkowski space is unitary for those values of the non-integer
power. The aim in this work is to add a $\phi^{4}$ interaction and
study the renormalization and critical properties of the resulting
non-local theory\footnote{This theory can also be obtained as the analytic regularized\cite{Bollini1964}
version of the usual $\phi^{4}$ local field theory.}. This study shows the relevance of this model in describing non-trivial
fixed points. The features and results of this work are summarized
as follows, 
\begin{itemize}
\item The theory to be considered is the free scalar theory studied in \cite{Trinchero:2012zn}
with the addition of a $\phi^{4}$ interaction term.
\item The contribution of the lowest order Feynman diagrams corresponding
to coupling constant renormalization, mass renormalization and field
renormalization are computed. This computation exemplifies general
issues about the renormalization of non-local field theories. The
procedure employed involves features that do not appear in the local
case. 
\item The previous calculation allows to compute the fixed point value for
the coupling constant and the critical exponents $\nu$ and $\eta$,
respectively. The corresponding results describe a theory which shows
asymptotic freedom in the UV and a non-trivial infrared fixed point
at finite coupling. The corresponding theory does not fulfill the
condition of reflection positivity.
\item In addition asuming the usual $n$-dimensional conformal algebra to
be a symmetry of the theory, the unitarity bounds are studied. They
agree with the ones obtained by requiring the condition of reflection
positivity.
\end{itemize}

\section{The action}

The free part of the action to be considered is essentially the same
as in \cite{Trinchero:2012zn} for\footnote{No infrared regulator is required for the following computations.}
$M=0$. The interaction part is $\phi{{}^4}$. In terms of the scalar
product of form fields mentioned above and described in \cite{Trinchero:2012zn},
the action is given by,
\begin{equation}
S=S_{0}+S_{I},\:\:S_{0}=\frac{1}{2}<d\phi,d\phi>,\:\:S_{I}=\frac{\lambda_{0}}{4!}<\phi^{2},\phi^{2}>\label{eq:action}
\end{equation}
evaluating the scalar products appearing in the last equation leads
to the following expression in terms of an integral over the integer
$n$-dimensional space,
\[
S_{0}=\int d^{n}x\;\frac{1}{2}\phi(-\square+m_{0}^{2})(-\square)^{-2\alpha}\phi,\;\;S_{I}=\frac{\lambda_{0}}{4!}\int d^{n}x\text{\ensuremath{\phi^{4}}}
\]
where, anticipating renormalization effects, a explicit mass term
has been included\footnote{This way of introducing a mass term is motivated by the calculation
of perturbative corrections appearing below. }. In what follows bare mass and coupling will be indicated by $m_{0}$
and $\lambda_{0}$ , the corresponding renormalized quantities will
be $m$ and $\lambda$. The Fourier transform of the free two point
function is therefore given by,
\[
<\phi\phi>(p)=\frac{1}{(p^{2}+m_{0}^{2})(p^{2})^{-2\alpha}}=\frac{\Gamma(1-2\alpha)}{\Gamma(-2\alpha)}\int_{0}^{1}da\,(1-a)^{-1-2\alpha}\frac{1}{\left(p^{2}+m_{0}^{2}a\right)^{1-2\alpha}}
\]
the second equality in the last equation is obtained using Feynman
parametrization. This last expression will be employed in the computations
below.

\section{Renormalization and the Critical exponents}

\subsection{Field renormalization\label{subsec:eta}}

Field renormalization is required at the two loop level. The corresponding
correction to the two point function is given by the following sunrise
diagram,
\noindent \begin{center}
\includegraphics[scale=0.2]{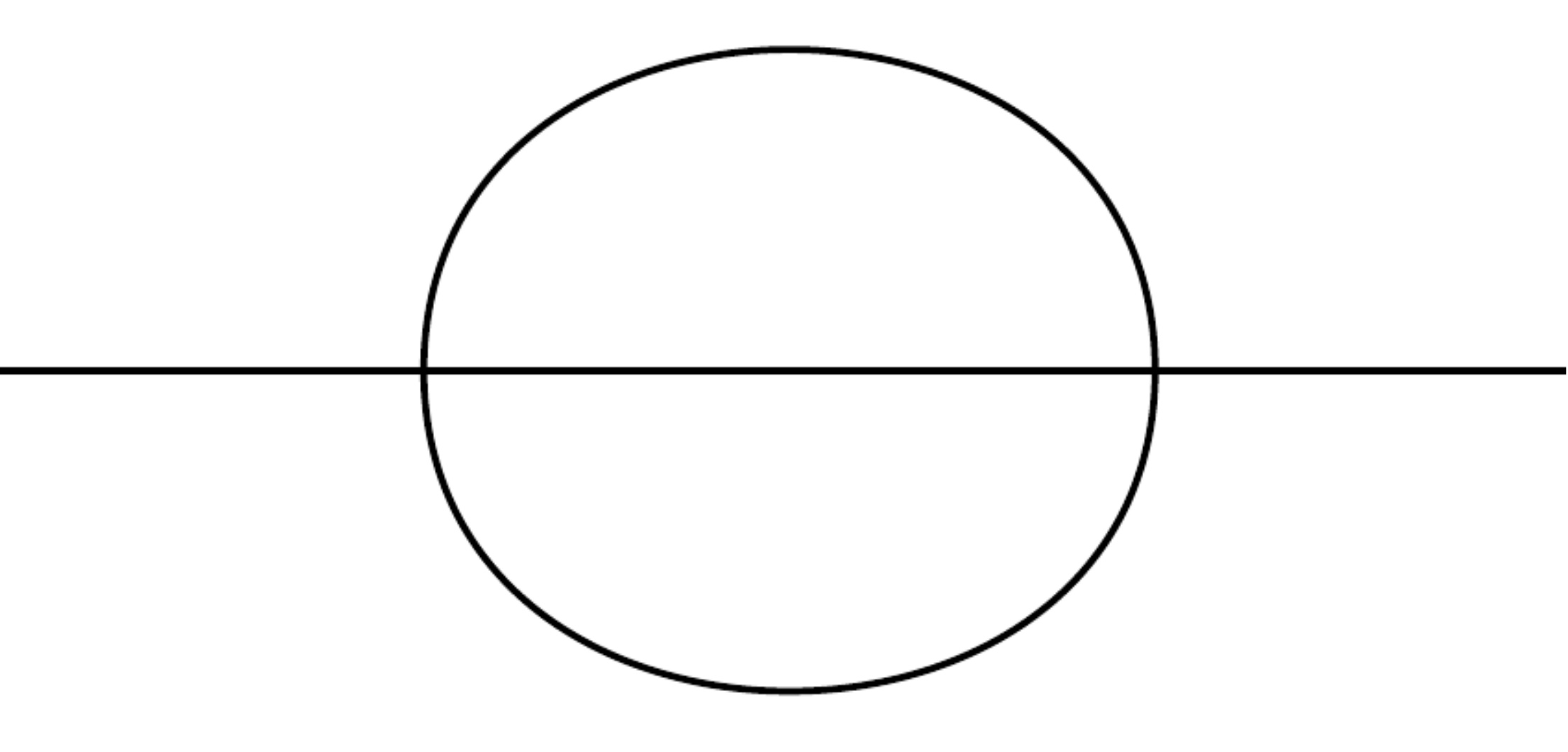}
\par\end{center}

\noindent The integral to be computed is,
\begin{align*}
a_{S}(p,\alpha) & =\left(\frac{\Gamma(1-2\alpha)}{\Gamma(-2\alpha)}\right)^{3}\int_{0}^{1}\left(\prod_{i=1}^{3}da_{i}\,(1-a_{i})^{-1-2\alpha}\right)I_{S}(p,\alpha,a_{1},a_{2},a_{3})\\
I_{S}(p,\alpha,a_{1},a_{2},a_{3}) & =\lambda_{0}^{2}\intop\frac{d^{n}q_{1}}{(2\pi)^{n}}\frac{d^{n}q_{2}}{(2\pi)^{n}}\frac{1}{(q_{1}^{2}+a_{1}m_{0}^{2})^{1-2\alpha}(q_{2}^{2}+a_{2}m_{0}^{2})^{1-2\alpha}[(p+q_{1}+q_{2})^{2}+a_{3}m_{0}^{2}]^{1-2\alpha}}
\end{align*}
For the purposes of this work it is convenient to expand the integrand
as a power series in $m^{2}.$ This leads to the following expression,
\[
I_{S}(p,\alpha,a_{1},a_{2},a_{3})=I_{0}(p,\alpha)+m_{0}^{2}(a_{1}+a_{2}+a_{3})I_{2}(p,\alpha)+\mathcal{O}(m_{0}^{4})
\]
where,
\begin{align*}
I_{0}(p,\alpha) & =\lambda_{0}^{2}\intop\frac{d^{n}q_{1}}{(2\pi)^{n}}\frac{d^{n}q_{2}}{(2\pi)^{n}}\frac{1}{(q_{1}^{2})^{1-2\alpha}(q_{2}^{2})^{1-2\alpha}[(p+q_{1}+q_{2})^{2}]^{1-2\alpha}}\\
I_{2}(p,\alpha) & =3\lambda_{0}^{2}\intop\frac{d^{n}q_{1}}{(2\pi)^{n}}\frac{d^{n}q_{2}}{(2\pi)^{n}}\frac{1}{(q_{1}^{2})^{1-2\alpha}(q_{2}^{2})^{1-2\alpha}[(p+q_{1}+q_{2})^{2}]^{2-2\alpha}}
\end{align*}
, introducing Feynman parametrizations to rewrite the integrands,
performing the momentum integrals and the integrals on the Feynman
parameters and taking $n=4$ leads to,
\begin{align*}
I_{0}(p,\alpha) & =p^{2(6\alpha+1)}\frac{4^{-2\alpha-3}\lambda_{0}^{2}\csc(4\pi\alpha)\Gamma(-6\alpha-1)\Gamma\left(\frac{3}{2}-2\alpha\right)B_{1}(4\alpha+2,2\alpha+1)}{\pi^{7/2}\Gamma(3-8\alpha)\Gamma(1-2\alpha)^{2}\Gamma(4\alpha)}\\
I_{2}(p,\alpha) & =\left(p^{2}\right)^{6\alpha}\frac{\lambda_{0}^{2}\Gamma(-6\alpha)\Gamma(2\alpha)\Gamma(2\alpha+1)^{2}}{256\pi^{4}\Gamma(1-2\alpha)^{2}\Gamma(2-2\alpha)\Gamma(6\alpha+2)}
\end{align*}
The coupling $\lambda_{0}$ has dimension $4-n-8\alpha$ in momentum
units, therefore for $n=4$ , it can be written as follows in terms
of an adimensional coupling $g_{0}$ as follows,
\begin{equation}
\lambda_{0}=g_{0}\,\mu^{-8\alpha}\Rightarrow g=\lambda\mu^{8\alpha}\label{eq:adcou-1}
\end{equation}
 noting that,
\begin{align*}
\left(\frac{\Gamma(1-2\alpha)}{\Gamma(-2\alpha)}\right)^{3}\int_{0}^{1}\left(\prod_{i=1}^{3}da_{i}\,(1-a_{i})^{-1-2\alpha}\right)(a_{1}+a_{2}+a_{3}) & =\frac{3}{1-2\alpha}\\
\left(\frac{\Gamma(1-2\alpha)}{\Gamma(-2\alpha)}\right)^{3}\int_{0}^{1}\left(\prod_{i=1}^{3}da_{i}\,(1-a_{i})^{-1-2\alpha}\right) & =1
\end{align*}
leads to,
\begin{align*}
a_{S}(p,\alpha) & =\frac{g_{0}^{2}}{(4\pi)^{4}}\mu^{-16\alpha}\left(p^{2}\right)^{6\alpha}\left[\frac{p^{2}}{12\alpha}+m_{0}^{2}\left(-\frac{1}{4\alpha^{2}}+\frac{1}{2\alpha}\right)+\mathcal{O}(\alpha^{0})\right]\\
 & =\frac{g_{0}^{2}}{(4\pi)^{4}}\mu^{-16\alpha}\frac{\left(p^{2}\right)^{6\alpha}}{12\alpha}\left[p^{2}+m_{0}^{2}\left(-\frac{3}{\alpha}+6\right)+\mathcal{O}(\alpha)\right]
\end{align*}
this result shows that $I_{0}(p,\alpha)$ is the relevant integral
for the field renormalization and that $I_{2}(p,\alpha)$ contributes
to mass renormalization. At this stage a recurrent situation in the
renormalization of these non-local theories shows up. Similar to what
happens in dimensional regularization the correction provided by a
given diagram, in this case the sunrise diagram, is proportional to
a power of the momentum which is not in general the same as the one
that originally appears in the Lagrandian. The integral $I_{0}(p,\alpha)$
gives a contribution\footnote{In dimensional regularization of the usual local $\phi^{4}$ theory
, the correction provided by the sunrise diagram is proportional to
$p^{2(d-3)}=p^{2(1-\epsilon)}$.} proportional to $p^{2(1+6\alpha)}$, while the original Lagrangian
has the power $p^{2(1-2\alpha)}$. The choice of the power of $p^{2}$
that appears in the kinetic term of the renormalized Lagrangian fixes
the finite contribution of this diagram. In other words if a different
power of $p^{2}$ is choosen than the finite contribution of the diagram
will also be different. The following way of rewritting $a_{S}(p,\alpha)$
illustrates this point, 
\begin{align}
a_{S}(p,\alpha) & =\frac{g_{0}^{2}}{(4\pi)^{4}}\mu^{-16\alpha)}\frac{\left(p^{2}\right)^{-2\alpha}}{12\alpha}\left(p^{2}\right)^{8\alpha}\left[p^{2}+m_{0}^{2}\left(-\frac{3}{\alpha}+6\right)+\mathcal{O}(\alpha)\right]\nonumber \\
 & =\frac{g_{0}^{2}}{(4\pi)^{4}}\frac{\left(p^{2}\right)^{-2\alpha}}{12\alpha}\left[p^{2}+m_{0}^{2}\left(-\frac{3}{\alpha}+6\right)+\mathcal{O}(\alpha)\right]\left(1+8\alpha\log\left(\frac{p^{2}}{\mu^{2}}\right)+\mathcal{O}(\alpha^{2})\right)\nonumber \\
 & =\frac{g_{0}^{2}}{(4\pi)^{4}}\frac{\left(p^{2}\right)^{-2\alpha}}{12\alpha}\left[p^{2}+m_{0}^{2}\left(-\frac{3}{\alpha}+6\right)+\mathcal{O}(\alpha)\right]\label{eq:as}
\end{align}
For $\alpha\to0$ the pole term of the last expression multiplied
by the symmetry\cite{Kleinert:2001ax} factor $\frac{1}{6}$ is the
one to be substracted. It is given by,
\[
\left(\frac{1}{6}a_{S}(p,\alpha\right)_{pole}=\frac{g_{0}^{2}}{(4\pi)^{4}}\frac{p^{2(1-2\alpha)}}{72\alpha}
\]
which leads to the renormalization constant,
\[
Z_{\phi}=1+\frac{g_{0}^{2}}{(4\pi)^{4}}\frac{1}{72\alpha}
\]
the function $\gamma$ is defined and given by,
\begin{align}
\gamma(g) & =\mu\left.\frac{\partial}{\partial\mu}\log Z_{\phi}^{\frac{1}{2}}\right|_{\lambda\,fixed}=\frac{1}{2}\frac{\partial}{\partial\log\mu}\log\left(1+\frac{\lambda_{0}^{2}\mu^{16\alpha}}{(4\pi)^{4}}\frac{1}{72\alpha}\right)\nonumber \\
 & =\frac{1}{2\,Z_{\phi}}\left(\frac{\lambda_{0}^{2}16\alpha}{(4\pi)^{4}72\alpha}\mu^{16\alpha}\right)=\frac{g^{2}}{9(4\pi)^{4}}+\mathcal{O}(g^{4})\label{eq:gamma}
\end{align}

\subsection{The fixed point and coupling constant renormalization}

The diagram to be considered is the one corresponding to the one loop
correction to the quartic coupling, i.e.,
\begin{center}
\includegraphics[viewport=-4cm 0bp 781bp 416bp,scale=0.15]{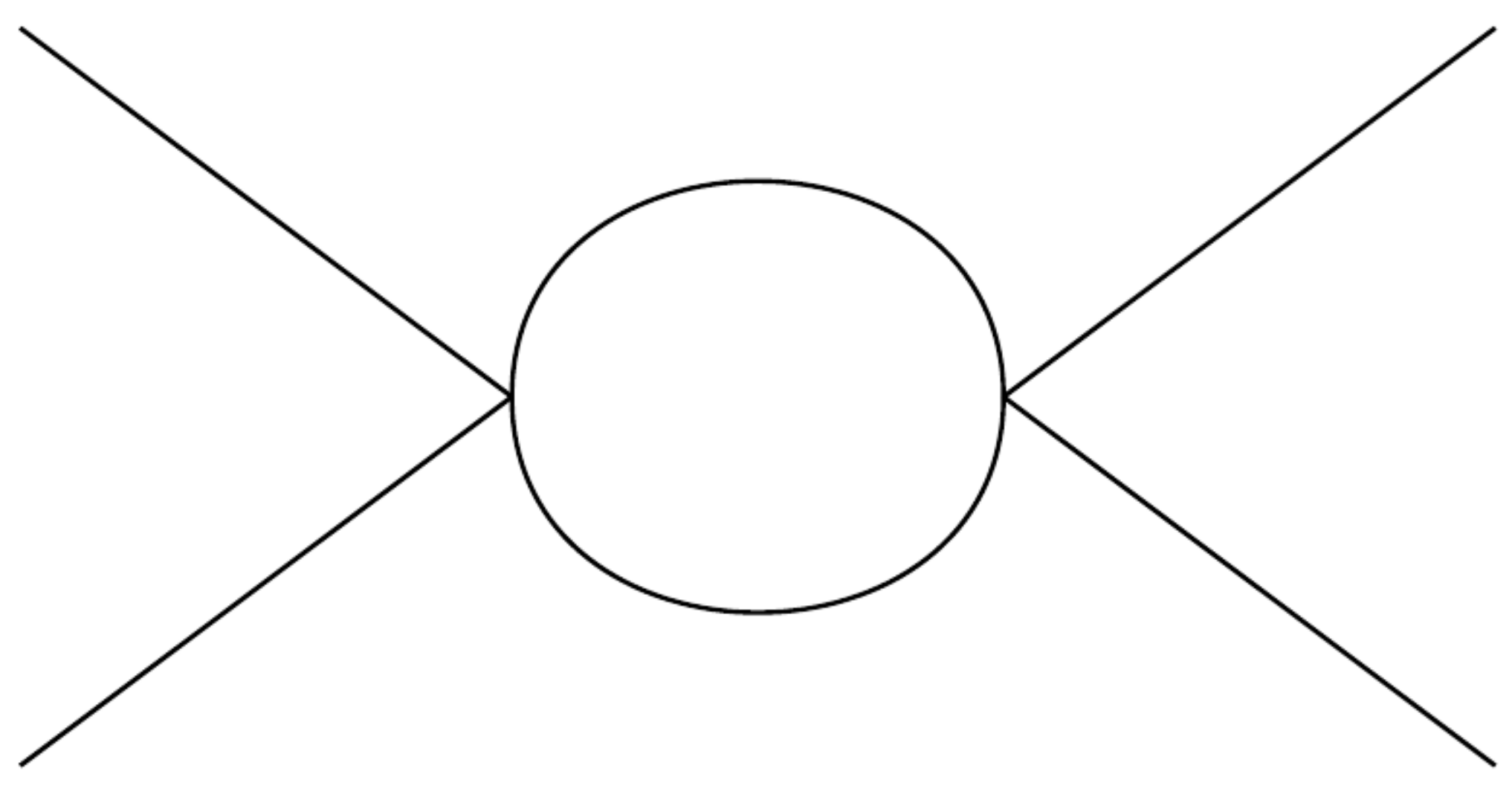} 
\par\end{center}

\noindent The integral to be computed is,
\begin{align*}
a_{F}(p,\alpha) & =\left(\frac{\Gamma(1-2\alpha)}{\Gamma(-2\alpha)}\right)^{2}\int_{0}^{1}\left(\prod_{i=1}^{2}da_{i}\,(1-a_{i})^{-1-2\alpha}\right)I_{F}(p,\alpha,a,b)\\
I_{F}(p,\alpha,a,b) & =\frac{3}{2}\lambda_{0}^{2}\int\,\frac{d^{n}q}{(2\pi)^{n}}\frac{1}{\left(q^{2}+m_{0}^{2}a\right)^{1-2\alpha}\left((p+q)^{2}+m_{0}^{2}b\right)^{1-2\alpha}}
\end{align*}
the factor $\frac{3}{2}$ coming from the $\frac{1}{2!}$ of the second
order term of the exponential and the contributions of $3$ diagrams
which give the same contribution. Introducing the Feynmann parametrization
and integraring over the $n$-moment $q$, leads to,
\begin{align*}
I_{F}(p,\alpha,a,b) & =\frac{3}{2}\frac{\lambda_{0}^{2}}{(4\pi)^{\frac{n}{2}}}\frac{\Gamma\left(2-\frac{n}{2}-4\alpha\right)}{\Gamma(1-2\alpha)^{2}}\int_{0}^{1}dx\frac{\left(m^{2}[ax+b(1-x)]+p^{2}(1-x)x\right)^{\frac{n}{2}-2+4\alpha}}{\left((1-x)x\right)^{2\alpha}}\\
 & \overset{_{n=4}}{=}\frac{3}{2}\frac{\lambda^{2}\mu^{8\alpha}}{(4\pi)^{2}}\frac{\Gamma\left(-4\alpha\right)}{\Gamma(1-2\alpha)^{2}}\int_{0}^{1}dx\left[\frac{\left(m_{0}^{2}+p^{2}(1-x)x\right)^{2}}{\mu^{4}(1-x)x}\right]^{2\alpha}\\
 & \overset{_{\alpha\ll1}}{=}\frac{3(\lambda_{0}\mu^{4\alpha})^{2}}{4(4\pi)^{2}}\left[-\frac{1}{2\alpha}+\int_{0}^{1}dx\log\left(\frac{\mu^{4}(1-x)x}{\left(m_{0}^{2}-p^{2}(x-1)x\right)^{2}}\right)\right]
\end{align*}
where in the second equality a parameter $\mu$ with dimensions of
mass has been introduced in order to make adimensional the argument
of the logarithm. In addition in the last equality only terms up to
$\mathcal{O}(\alpha^{0})$ has been kept. In the minimal substraction
scheme only the first term in the square bracket of the last expression
will be relevant in defining the renormalized coupling $\lambda_{R}$.
This term is independent of $a$ and $b$, therefore,
\begin{align*}
a_{F}(p,\alpha) & =\left(\frac{\Gamma(1-2\alpha)}{\Gamma(-2\alpha)}\right)^{2}\int_{0}^{1}da\,db(1-a)^{-1-2\alpha}(1-b)^{-1-2\alpha}\frac{3(\lambda_{0}\,\mu^{4\alpha})^{2}}{4(4\pi)^{2}}\left(-\frac{1}{2\alpha}\right)\\
 & =\frac{3(\lambda_{0}\mu^{4\alpha})^{2}}{4(4\pi)^{2}}\left(-\frac{1}{2\alpha}\right)
\end{align*}
Taking into account the computation in the last subsection, this leads
to the following renormalizaed coupling,
\[
\lambda=\frac{\lambda_{0}Z_{\phi}^{2}}{Z_{g}},\;\;Z_{g}=1+\frac{3}{4}\frac{\lambda_{0}}{(4\pi)^{2}}\frac{\left(\mu^{2}\right)^{4\alpha}}{(-2\alpha)}+\mathcal{O}(\lambda_{0}^{2})\,\,,Z_{\phi}=1+\frac{g_{0}^{2}}{(4\pi)^{4}}\frac{1}{72\alpha}
\]
the beta function corresponding to the renormalized adimensional coupling
$g$ fulfills,
\[
\beta(g_{R})=\mu\frac{d}{d\mu}g_{R}=\mu\frac{d}{d\mu}\left(\frac{\lambda Z_{\phi}^{2}}{Z_{g}}\mu^{8\alpha}\right)=8\alpha g_{R}+2g_{R}Z_{\phi}^{-1}\mu\frac{d}{d\mu}Z_{\phi}+g_{R}Z_{g}^{-1}\mu\frac{d}{d\mu}Z_{g}=8\alpha g_{R}+4g_{R}\gamma+g_{R}\frac{3}{8\alpha}\frac{1}{(4\pi)^{2}}\beta(g_{R})
\]
which implies,
\[
\beta(g_{R})=8\alpha g_{R}+\frac{6g_{R}^{2}}{2(4\pi)^{2}}+4\frac{g_{R}^{3}}{9(4\pi)^{4}}
\]
Neglecting negative values of $g,$ which make the theory unstable,
the figure below shows a plot of this function for $\alpha=\pm0.01$,
\noindent \begin{center}
\includegraphics{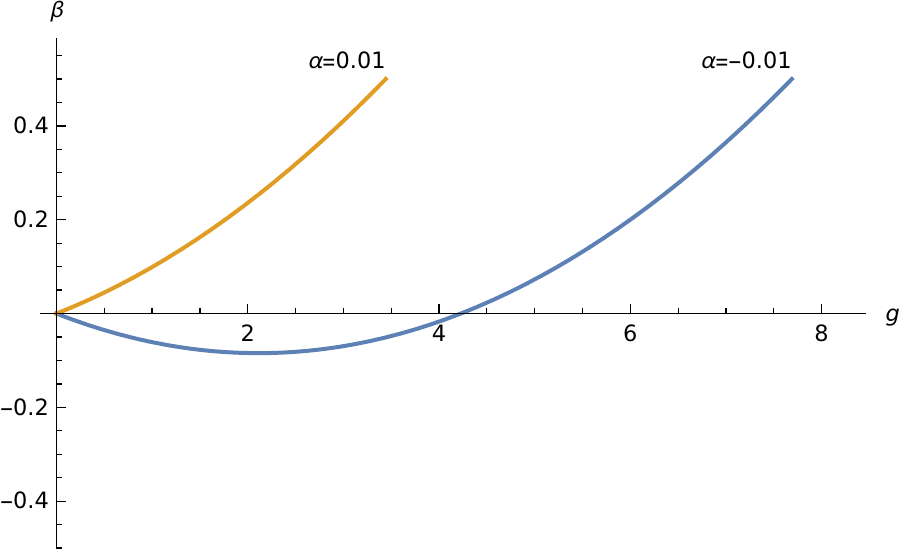}
\par\end{center}

\noindent This figure shows that for $\alpha>0$ the theory has asymptotic
freedom in the infrared. However for $\alpha<0$ this $4$-dimensional
theory presents asymptotic freedom(AF) in the ultraviolet(UV). This
a remarkable result since it is usually believed that non-abelian
gauge bosons are required in order to get AS in the UV. However as
the analysis in \cite{Trinchero:2017weu} shows, the theory for $\alpha<0$
does not satisfy the requirement of reflection positivity(RP). This
means that the Wick rotated theory in Minkowski space does not provide
a unitary representation of the Poincarè group, which implies that
no unitary evolution can be defined in this space. Alternatively,
as will be shown in the next section, the unitarity bounds are violated
for $\alpha<0$. This does not mean that the Euclidean theory is useless,
indeed many useful statistical mechanical models fail to satisfy RP. 

The fixed point $g^{\star}$ is defined by $\beta(g^{\star})=0$ .
Writting the solution of this last equation as a power series in $\alpha$,
\[
g^{\star}=g_{0}+g_{1}\alpha+g_{2}\alpha^{2}+\cdots
\]
leads to two solutions, the Gaussian fixed point $g^{\star}=0$ and,
\[
g^{\star}=-\frac{8}{3}(4\pi)^{2}\alpha-\frac{256}{243}(4\pi)^{2}\alpha^{2}
\]

\subsection{Mass renormalization}

The one loop correction to the two point function is given by the
following diagram,
\begin{center}
\includegraphics[scale=0.4]{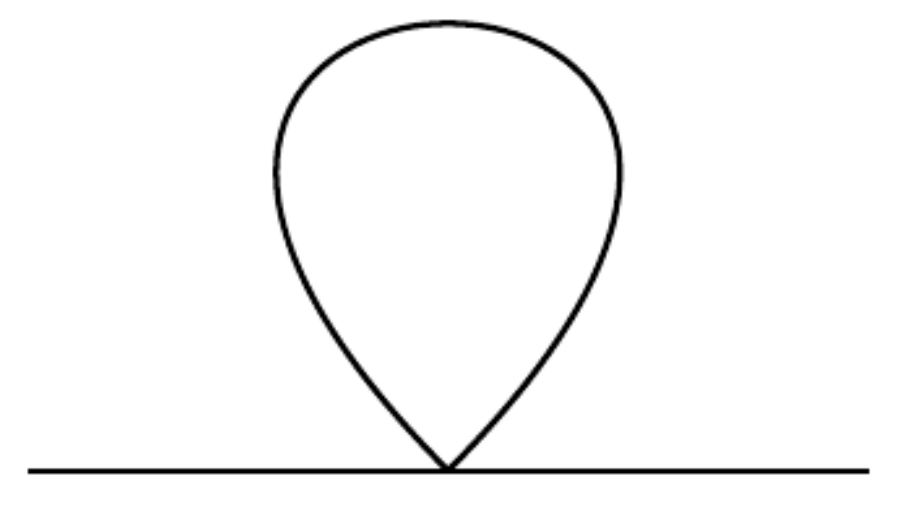}
\par\end{center}

\noindent The integral to be computed is,
\begin{align}
a_{T}(\alpha) & =\frac{\Gamma(1-2\alpha)}{\Gamma(-2\alpha)}\int_{0}^{1}da\,(1-a)^{-1-2\alpha}I_{T}(\alpha,a)\label{eq:ait}\\
I_{T}(\alpha,a) & =-\lambda_{0}\intop\frac{d^{n}q}{(2\pi)^{n}}\frac{1}{(q^{2}+m_{0}^{2}a)^{1-2\alpha}}\nonumber 
\end{align}
leading to,
\begin{align*}
I_{T}(\alpha,a) & =-\frac{\lambda_{0}}{(4\pi)^{\frac{n}{2}}}\frac{\Gamma(1-\frac{n}{2}-2\alpha)}{\Gamma(1-2\alpha)}(m_{0}^{2}a)^{\frac{n}{2}-1+2\alpha}
\end{align*}
replacing in (\ref{eq:ait}) leads to,
\begin{align*}
a_{T}(\alpha) & =\frac{\Gamma(1-2\alpha)}{\Gamma(-2\alpha)}\int_{0}^{1}da\,(1-a)^{-1-2\alpha}\left(-\frac{\lambda_{0}}{(4\pi)^{\frac{n}{2}}}\frac{\Gamma(1-\frac{n}{2}-2\alpha)}{\Gamma(1-2\alpha)}(m_{0}^{2}a)^{\frac{n}{2}-1+2\alpha}\right)\\
 & \overset{_{n=4}}{=}\frac{-g_{0}\mu^{-8\alpha}}{16\pi\,\sin(2\pi\alpha)}(m_{0}^{2})^{1+2\alpha}=\frac{-g_{0}}{16\pi\,\sin(2\pi\alpha)}(m_{0}^{2})^{1-2\alpha}\left(\frac{m_{0}^{2}}{\mu^{2}}\right)^{4\alpha}\\
 & =(m_{0}^{2})^{1-2\alpha}\left(\frac{-g_{0}}{(4\pi)^{2}2\alpha}+\mathcal{O}(\alpha^{0})\right)
\end{align*}
where in the second equality the dimensional coupling $\lambda$ has
been expressed in terms of the adimensional coupling $g$ by means
of (\ref{eq:adcou-1}). As was mentioned for the case of the sunrise
diagram, in this case also the power of $p^{2}$ appearing in the
correction is different from the one appearing in the Lagrangian.
In a similar way as for the sunrise, the choice of the power to appear
in the renormalized Lagrangian fixes the finite contribution of this
diagram. This last point is illustrated by the following computation
of the correction to the proper two point function\footnote{It is worth noting that if the mass were included with a kinetic term
of the form, $\mathcal{L}_{0}=\phi(-\square+m^{2})^{1-2\alpha}\phi$
then the singular contribution of this diagram when $\alpha\to0$
could not be absorved by mass renormalization, in other words the
counterterm required to cancel the divergence when $\alpha\to0$ would
not be of the form $\mathcal{L}_{0}$.

},
\begin{align*}
\Gamma_{2}(p) & =(p^{2}+m_{0}^{2})(p^{2})^{-2\alpha}-\frac{1}{2}a_{T}(\alpha)\\
 & =(p^{2}+m_{0}^{2})(p^{2})^{-2\alpha}+(m_{0}^{2})^{1-2\alpha}\frac{g_{0}}{(4\pi)^{2}4\alpha}+C\\
 & =(p^{2}+m_{0}^{2})(p^{2})^{-2\alpha}+m_{0}^{2}(p^{2})^{-2\alpha}\left(\frac{m_{0}^{2}}{p^{2}}\right)^{-2\alpha}\frac{g_{0}}{(4\pi)^{2}4\alpha}+C\\
 & =(p^{2}+m_{0}^{2}\left(1+\frac{g_{0}}{(4\pi)^{2}4\alpha}\right))(p^{2})^{-2\alpha}+C'
\end{align*}
where $C$ and $C'$, denote terms that converge for $\alpha\to0$.
Therefore, taking into account the computation in subsection \ref{subsec:eta},
the renormalized mass $m$ in the minimal substraction scheme is given
by,
\begin{align*}
m^{2} & =m_{0}^{2}\frac{Z_{\phi}}{Z_{m^{2}}}
\end{align*}
where, up to $\mathcal{O}(g^{2})$,
\[
Z_{m^{2}}=1-\frac{g_{0}}{(4\pi)^{\frac{n}{2}}}\frac{1}{4\alpha}
\]
the beta function $\gamma_{m}$ for the mass is given by,
\begin{align*}
\gamma_{m}(g) & =\frac{\mu}{m}\,\frac{\partial m}{\partial\mu}=\frac{1}{2}\left(\frac{\mu}{Z_{\phi}}\frac{\partial\log Z_{\phi}}{\partial\mu}-\mu\frac{\partial Z_{m^{2}}}{\partial\mu}\right)=\gamma+\frac{1}{2}\frac{\beta(g)}{(4\pi)^{\frac{n}{2}}4\alpha}\\
 & =\frac{1}{2}\beta(g)\left(\frac{1}{(4\pi)^{\frac{n}{2}}4\alpha}+\frac{2g_{0}}{(4\pi)^{4}72\alpha}\right)=\frac{g}{(4\pi)^{2}}+\frac{g^{2}}{9(4\pi)^{4}}+\mathcal{O}(g^{3})
\end{align*}

\subsection{The critical exponents $\nu$ and $\eta$}

\noindent These critical exponents are related to the fixed point
values $\gamma^{\star}$ and $\gamma_{m}^{\star}$ of the functions
$\gamma$ and $\gamma_{m}$. They are given by,
\[
\nu=\frac{1}{2-2\gamma_{m}^{\star}}\;\;\;,\eta=2\gamma^{\star}
\]
 The non-trivial fixed point is given by,
\[
g^{\star}=-\frac{8}{3}(4\pi)^{2}\alpha-\frac{256}{243}(4\pi)^{2}\alpha^{2}
\]
 the fixed point values $\gamma_{m}^{\star}$ and $\gamma^{\star}$
are therefore given by,
\begin{align*}
\gamma_{m}^{\star} & =\gamma_{m}(g^{\star})=-\frac{8\alpha}{3}+\frac{256}{243}\left(12\pi^{2}-1\right)\alpha^{2}+\frac{65536\pi^{2}\alpha^{3}}{6561}+\mathcal{O}(\alpha^{3})\\
\gamma^{\star} & =\gamma(g^{\star})=\frac{64\alpha^{2}}{81}+\frac{4096\alpha^{3}}{6561}
\end{align*}
which imply,
\begin{align}
\nu & =\frac{1}{2}-\frac{4\alpha}{3}+\frac{32}{243}\left(23+48\pi^{2}\right)\alpha^{2}-\frac{256\left(171+736\pi^{2}\right)\alpha^{3}}{6561}\nonumber \\
\eta & =2\left(\frac{64\alpha^{2}}{81}+\frac{4096\alpha^{3}}{6561}\right)\label{eq:eta}
\end{align}

\noindent It is worth noting that the value of $\alpha$ is related
to the dimension of space. A free propagator at the Gaussian fixed
point in $4$-dimensions, should behave as $\frac{1}{|x|^{2}}$, this
corresponds to small values of $\alpha$, as the ones employed in
the last figure. The critical exponents for the non-Gaussian fixed
point for $\alpha=-0.01$ are,
\begin{align*}
\nu\overset{_{\alpha=-0.01}}{=}0.52\;\;\; & ,\eta\overset{_{\alpha=-0.01}}{=}0.0001
\end{align*}

\noindent Following the same reasoning in $3$ dimensions, the $\frac{1}{|x|}$
behaviour of the free propagator givesa $\alpha=-\frac{1}{4}$ .This
is the value of $\alpha$ which corresponds to the $\epsilon=1$ in
the $\epsilon$-expansion. In the same spirit as in the case of the
$\epsilon$-expansion, the critical exponents for the non-Gaussian
fixed point can be computed for this last value of $\alpha$. Replacing
$\alpha=-\frac{1}{4}$ in (\ref{eq:eta}) leads to the following values
for the critical exponents,
\[
\nu\overset{_{\alpha=-\frac{1}{4}}}{=}4.92\;\;\;,\eta\overset{_{\alpha=-\frac{1}{4}}}{=}0.079
\]
which, in comparison with the values obtained with the $\epsilon$-expansion,
signficantly differs form the $3d$-Ising model critical exponents.
This shows that this fixed point does not describe the $3d$-Ising
model critical point.

\section{Relation with $\epsilon$ expansion}

For each diagram there is a way to obtain its divergent contribution(when
$\alpha\to0$) from the corresponding one in the $\epsilon$ expansion.
In order to show this let us consider the superficial degree of divergence(SDD)
for both theories, the one considered in this paper described by the
action (\ref{eq:action}), from now on the $\alpha$-theory and the
usual $\phi^{4}$ theory dimensionaly regularized to a dimension $d=4-\epsilon$,
from now on the $\epsilon$-theory. The SDD for a proper graph $G$
in the $\epsilon$-theory is given by,
\[
\omega_{\epsilon}(G)=4-\epsilon(1+V)+\left(\frac{\epsilon}{2}-1\right)E
\]
where $V$ denotes the number of vertices and $E$ the number of external
legs. For the case of the $\alpha$-theory the SDD can be computed
to give,
\[
\omega_{\alpha}(G)=4+8\alpha V-(1+2\alpha)E
\]
which of course coincide for $\alpha=\epsilon=0$. Note that there
is no replacement of $\epsilon$ as a function of $\alpha$ such that
for any $V$ and $E$ the following equality holds\footnote{If such a replacement where posible then an expansion in powers of
$\alpha$ would be the same as the $\epsilon$ expansion.},
\[
\omega_{\epsilon(\alpha)}(G)=\omega_{\alpha}(G)
\]
However for each given $V$ and $E$ there is a replacement. This
is shown in the the table below, which compares the SDD and the renormalization
constants for the diagrams considered in the previous section, 
\[
Z(\epsilon)\to Z(\alpha)
\]
\vspace{0.7cm}

\noindent %
\begin{tabular}{|c|c|c|c|c|c|}
\hline 
Diagram & $\omega_{\epsilon}$ & $\omega_{\alpha}$ & $Z(\epsilon)$ & $Z(\alpha)$ & $\begin{array}{ccc}
\omega_{\epsilon} & \to & \omega_{\epsilon}\\
Z(\epsilon) & \to & Z(\alpha)
\end{array}$\tabularnewline
\hline 
\hline 
\,\,\,\,\includegraphics[viewport=20bp -20bp 367bp 179bp,clip,scale=0.1]{2-point.pdf} & $2-\epsilon$ & $2+4\alpha$ & $Z_{m^{2}}(\epsilon)=1+\frac{g}{(4\pi)^{2}\epsilon}$ & $Z_{m^{2}}(\alpha)=$1-$\frac{g}{(4\pi)^{2}4\alpha}$ & $\epsilon\to-4\alpha$\tabularnewline
\hline 
\,\,\,\,\includegraphics[viewport=20bp -30bp 801bp 386bp,clip,scale=0.04]{fi4.pdf} & $-\epsilon$ & $8\alpha$ & $Z_{g}(\epsilon)=1+\frac{3g}{(4\pi)^{2}\epsilon}$ & $Z_{g}(\alpha)=1-\frac{3g}{(4\pi)^{2}8\alpha}$ & $\epsilon\to-8\alpha$\tabularnewline
\hline 
\,\,\includegraphics[viewport=20bp -30bp 869bp 377bp,clip,scale=0.04]{sunrise.pdf} & $2-2\epsilon$ & $2+12\alpha$ & $Z_{\phi}(\epsilon)=1-\frac{g^{2}}{(4\pi)^{4}12\epsilon}$ & $Z_{\phi}(\alpha)=1+\frac{g^{2}}{(4\pi)^{4}72\alpha}$ & $\epsilon\to-6\alpha$\tabularnewline
\hline 
\end{tabular}

\vspace{0.5cm}
This table shows that knowing the SDD of a given diagram in both theories
allows to obtain the renormalization constant in one theory knowing
the renormalization constant in the other. In other words for a given
diagram $G$ the same replacement that sends $\omega_{\epsilon}(G)$
to $\omega_{\alpha}(G)$, sends $Z_{G}(\epsilon)$ to $Z_{G}(\alpha)$.
This fact shows that an expansion in powers of $\alpha$ and the $\epsilon$
expansion, are not same and describe different critical theories,
this is so because of the non-trivial dependence of this replacement
on the diagram considered.

\section{Unitarity bounds}

\subsection{The conformal algebra in $n$-dimensions}

The action (\ref{eq:action}) is invariant under conformal transformations\footnote{Given that this theory can be thought as a theory depending on derivatives
of the field of any order, then there should be an nfinite number
of conserved charges. This assertion is not analysed in this paper. }. It is assumed that there exists conserved charges implementing these
transformations at the level of the field. The conformal algebra for
dimensions $n\geq3$ is given by,

\begin{align*}
[D,P_{\mu}] & =iP_{\mu}\\{}
[P_{\rho},L_{\mu\nu}] & =i(\eta_{\rho\mu}P_{\nu}\text{\textminus}\eta_{\rho\nu}P_{\mu})\\{}
[D,K_{\mu}] & =\text{\textminus}iK_{\mu}\\{}
[K_{\mu},P_{\nu}] & =2i(\eta_{\mu\nu}D\text{\textminus}L_{\mu\nu})\\{}
[K_{\rho},L_{\mu\nu}] & =i(\eta_{\rho\mu}K_{\nu}\text{\textminus}\eta_{\rho\nu}K_{\mu})\\{}
[L_{\mu\nu},L_{\rho\sigma}] & =i(\eta_{\nu\rho}L_{\mu\sigma}+\eta_{\mu\sigma}L_{\nu\rho}\text{\textminus}\eta_{\mu\rho}L_{\nu\sigma}\text{\textminus}\eta_{\nu\sigma}L_{\mu\rho})
\end{align*}
where $P_{\mu}$ are the generators of traslations, $L_{\mu\nu}$
the generators of rotations in the $\mu-\nu$ plane, $D$ the generator
of dilatations and $K_{\mu}$ the generators of special conformal
transformations. In cylindrical coordinates the hermiticity properies
of operators are such that\cite{Qualls:2015qjb},
\[
P_{\mu}^{\dagger}=K_{\mu}.
\]

\subsection{Positive definite inner products and bounds for $\alpha$}

For a spinless primary state $|\Delta>$ the commutation relation
between $P_{\mu}$ and $K_{\nu}$ can be used to show that,
\begin{align}
\left|P_{\mu}|\Delta>\right|^{2} & >0\:\:\:\Rightarrow\Delta>0\nonumber \\
\left|P_{\mu}P_{\nu}|\Delta>\right|^{2} & >0\:\:\:\Rightarrow\Delta>\frac{n-2}{2}\label{eq:unitb}
\end{align}
 for a space of dimension $n$. For the free theory $\lambda=0$,
the dimension of the field $\phi$ is,

\[
[\phi]=\frac{n-2+4\alpha}{2}
\]
thus the unitarity bound (\ref{eq:unitb}) implies,
\[
\alpha>0
\]
For the interacting theory,
\[
[\phi]=\frac{n-2+4\alpha}{2}+\frac{\eta}{2}
\]
thus the unitarity bound implies,
\[
\alpha>-\frac{\eta}{4}
\]
therefore, using (\ref{eq:eta}), this implies that,
\[
\alpha+\frac{1}{2}\left(\frac{64\alpha^{2}}{81}+\frac{4096\alpha^{3}}{6561}\right)>0
\]
the polynomial on the l.h.s. of the last inequality has only one real
root at $\alpha=0$, and the last inequality is equivalent to $\alpha>0$.
Showing that the free theory unitarity bound is stable under the corrections
computed in this work.

\section{Concluding remarks}

Conclusions and further research motivated by this work are summarized
in the series of remarks given below,
\begin{itemize}
\item It was shown that introducing a non-local kinetic term for a scalar
field with interaction $\phi^{4},$ it is possible to get asymptotic
freedom in the UV without including non-abelian vector fields. For
$\alpha<0$, the resulting theory can not be Wick rotated to obtain
a field theory over Minkowski space realizing a unitary representation
of the Poincarè group. This is so becuse the condition of reflection
positivity is not fulfilled for these values of $\alpha$.
\item In order to obtain that result the renormalization of the theory was
considered. This was done for the first corrections to the two and
four point functions. The concrete renormalization procedure shows
features different from the case of local theories, which all the
same make sense. In this respect allthough the usual renormalization
program in field theory is formulated for local interactions, nevertheless
the properties of a system involving non-local terms under rescaling
of distances is in the same footing as a system involving only local
ones from the point of view of the Wilsonian renormalization group.
\item The fact that reflection positivity does not hold for $\alpha<0$,
is confirmed by the violation for $\alpha<0$ of the unitarity bounds
obtained assuming that the theory provides a representation of the
conformal algebra. This rise up the question of the symmetries for
the non-local action (\ref{eq:action}), which is an interesting subject
to be considered.
\end{itemize}
Summarizing , it is believed that the study of non-local field theories
can enlarge our knowledge about the fixed points and renormalization
group flows in the space of all possible couplings mentioned in the
introduction.\vspace{0.5cm}

\textbf{Acknowledgements. }

I am deeply indebted to G. Torroba for sharing his expertise on the
renormalization group and for many enlightening discussions.

\bibliographystyle{unsrt}
\addcontentsline{toc}{section}{\refname}\bibliography{Bibliography,../../adsqcd/nf/Bibliography}

\end{document}